\documentclass[aps,prb,showpacs,twocolumn]{revtex4}

\usepackage{graphicx}
\usepackage{amsmath,amssymb,amsfonts}
\begin{document}

\title{Density-functional studies of spin-orbit splitting in graphene on
metals}

\author{Z. Y. Li$^{1}$}
\author{S. Qiao$^{1}$}
\author{Z. Q. Yang$^{1,2}$}
\email{zyang@fudan.edu.cn}
\author{R. Q. Wu$^{3}$}

\affiliation{$^{1}$State Key Laboratory of Surface Physics and
Department of Physics,
Fudan University, Shanghai 200433, China\\
$^{2}$Department of Chemistry, Northwestern University, Evanston, Illinois
60208, USA \\
$^{3}$Department of Physics and Astronomy, University of California, Irvine,
California 92697-4575, USA}

\date{\today}

\begin{abstract}
Spin-orbit splitting in graphene on Ni, Au, or Ag (111) substrates was
examined on the basis of density-functional theory. Graphene grown on the
three metals was found to have Rashba splitting of a few or several tens of
meV. The strong splitting obtained on Au or Ag substrates was mainly
ascribed to effective hybridization of graphene $p_{z}$ state with Au or Ag $%
d_{z^{2}}$ states, rather than charge transfer as previously proposed. Our
work provides theoretical understandings of the metal-induced Rashba effect
in graphene.
\end{abstract}


\keywords{}-
\maketitle

Graphene has attracted extensive attention in recent years due to its unique
and remarkable electronic properties, such as gapless-semiconductor band,
existence of pseudospin, and high electronic mobility at room temperature.%
\cite{geim,nov,zhang} These features are highly desirable for the
development of next-generation microelectronic and spintronic
devices.\cite{nature,sem} Spin currents in graphene can be
manipulated using various electronic tactics, in particular
through exchange and spin-orbit (SO) interactions,
which are now among the most active research topics in several realms.\cite%
{sem,datta,aws} The intrinsic SO effect in pure graphene layers is
nevertheless very weak, 0.1$\sim $0.37 meV in flat graphene sheet or carbon
nanotube \cite{kane,kue,yao}, insufficient for practical use due to the low
nuclear charge of the carbon atom. It was hence very exciting when Dedkov
\textit{et al} reported an extraordinarily large Rashba\cite{ra} SO
splitting (225 meV) for the $\pi $\ states of epitaxial graphene layers on
the Ni (111) substrate through their angle-resolved photoemission studies.%
\cite{ded} It appears that the SO effect or hybridization in
graphene can be tuned through effective electric field across the
interface. However, this result was challenged by Rader \textit{et
al} \cite{rad} who found that the sum of Rashba and exchange
splitting in the graphene layer on either Ni (111) or Co(0001) is
less 45 meV. They pointed out that the Rashba effect can be
strongly enhanced by intercalation of one monolayer of Au between
graphene and Ni (111). Clearly, to discriminate these
contradictory experimental results and furthermore to understand
the mechanism of substrate-induced SO splitting in graphene are
crucial for the progress of graphene physics.

In this Letter, we report results of density-functional theory (DFT) \cite%
{kohn} calculations for the electronic and magnetic properties of
graphene on Ni, Au, or Ag (111) films. Interestingly, the Rashba
splitting in graphene on Au and Ag can be significantly enhanced
by strong hybridization of graphene $p_{z}$ state and metal
$d_{z^{2}}$ state. The SO splitting is found to be almost
independent of the charge transfer between graphene and its
substrates, contradicting to effective electronic field model
proposed in Ref.\cite{ded}.

\begin{figure}[tbp]
\resizebox{7.0cm}{!}{\includegraphics*[48,580][475,702]{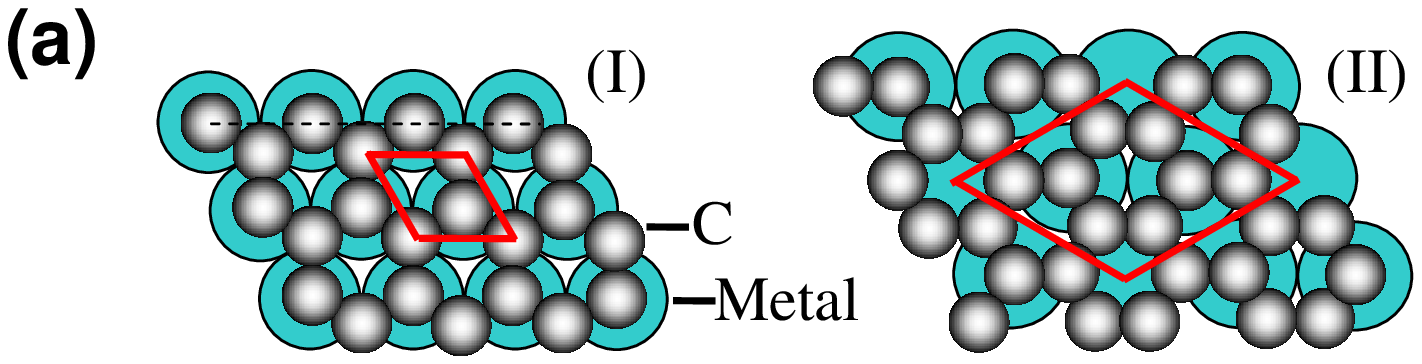}}
\resizebox{8.8cm}{!}{\includegraphics*[7,318][543,676]{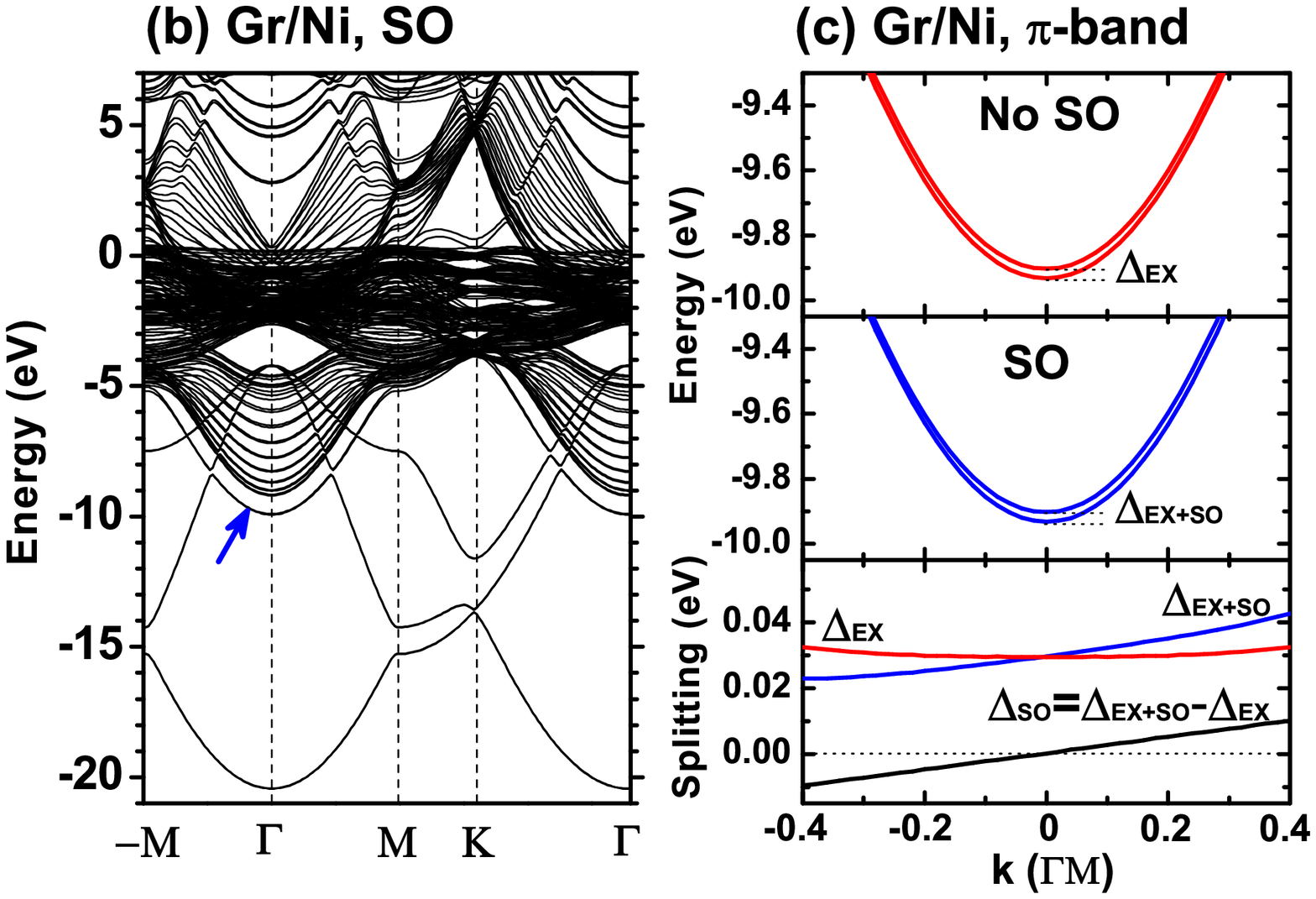}}
\caption{(Color online) (a) Two possible configurations of
graphene adsorbed on metal (111) substrates. The rhombus gives the
unit cell along the graphene plane. (b) The energy bands of Gr/Ni
in the configuration (I) with SO interaction. The blue arrow
indicates the graphene $\protect\pi $ bands around $\Gamma $. (c)
The enlarged $\protect\pi $ bands around $\Gamma $\ without or
with SO interactions. The exchange and/or SO splitting of the
bands are also given. The $k$ point in (c) is in the unit of the vector $%
\Gamma M$.}
\end{figure}

The electronic structures of graphene on metal (111) substrates,
abbreviated as Gr/M (M = Ni, Au, and Ag), were calculated by using
the VASP code at the level of local spin-density approximation
(LSDA).\cite{lsda} The projector-augmented wave (PAW)
pseudopotentials were employed to describe the effect of core
electrons. The equilibrium structures were obtained through
structural relaxation until the Hellmann-Feynman forces were less
than 0.05 eV/\r{A}.\cite{prb} The Gr/M systems were modeled by a
periodic slab geometry, with a vacuum of at least 10 \r{A}\
between two neighboring slabs. Each slab contains one graphene
layer and N atomic layers of metal. Two adsorption configurations
of graphene on the metal substrates were considered, as depicted
in Fig.1(a). We found that configuration (I) is more
stable for graphene on Ni (111), as it gives very small lattice mismatch ($\sim$%
1\%). For the same reason, configuration (II) is preferred for graphene on
the larger Au and Ag (111) lattices \cite{kar,gio}. As a benchmark
calculation for the SO effect, we first determined the SO splitting of the
surface states (SS) of the pure Au (111) film near the Fermi level (E$_{F}$,
set as energy zero).\cite{nico} Our result, $\Delta E_{SO}\sim $100 meV,
agrees well with data in Ref.\cite{nico}.

The band structures of Gr/Ni with N=13\cite{bertoni} are given in Fig.1(b)
and (c). The direction of Ni magnetization is set to be perpendicular to $%
\Gamma $M. Despite the strong perturbation from the Ni (111) substrate, one
can still easily trace several graphene bands in Fig.1(b), e.g., the
graphene $\sigma $ and $\pi $ states at -4 and -10 eV in the vicinity of the
$\Gamma $ point. Compared to the band structures of pure graphene, these
states are spin polarized, and shifted downward in energy by about 1.2 and
2.2 eV, respectively. Particularly, the feature conical points at $K$ near E$%
_{F}$ are destroyed in Fig.1(b), due to broken equivalence of \textit{A} and
\textit{B} sublattices through the interaction with Ni. These results are in
agreement with photoemission measurements\cite{gash}.

Now we zoom in to explore spin splitting and SO effect of the $\pi
$ states
of graphene along the -$M\Gamma M$ line, following the experimental work\cite%
{ded,rad,gierz}. To separate contributions from different factors, we
studied cases either with or without the SO interaction. As illustrated in
Fig.1(c), bands without the SO interaction are symmetric about the $\Gamma $
point and show an induced exchange splitting ($\Delta _{EX}$) of 30 meV on
the magnetic Ni substrate. After considering the SO interaction, the energy
splitting ($\Delta _{EX+SO}$) contains two parts: exchange and SO ($\Delta
_{SO}$). The value of $\Delta _{SO}$ can be extracted through $\Delta
_{SO}=\Delta _{EX+SO}-\Delta _{EX}$. The linear relationship of $\Delta _{SO}
$ versus \textit{k} in Fig.1(c) indicates that the SO interaction is indeed
the Rashba type ($\Delta _{SO}=2\alpha _{R}k$, where $\alpha _{R}$ is Rashba
strength).\cite{ra,rad} The Rashba splitting obtained from our calculations
is about 10 meV, in consistent with the experimental data in Ref. \cite{rad}%
. In comparison with the SO splitting (0.37 meV) in curved graphene\cite{kue}%
, the Ni-induced SO splitting in the graphene $\pi $ bands is relatively
larger.

\begin{figure}[tbp]
\resizebox{8.5cm}{!}{\includegraphics*{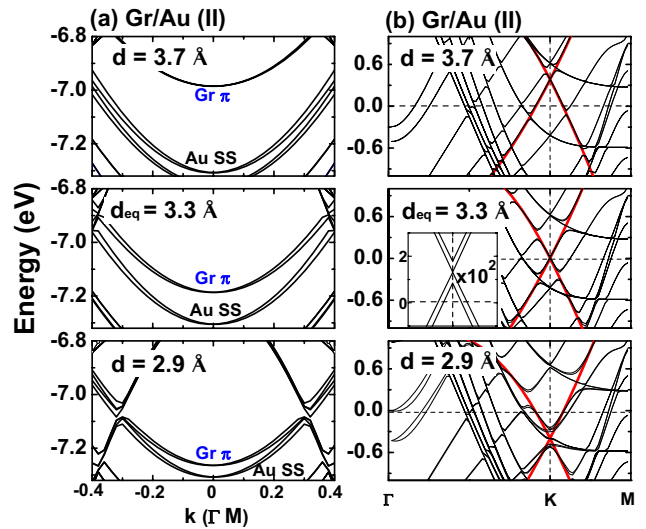}} \caption{(Color
online) The bands of Gr/Au in the configuration (II) with N=9 (a)
around -7.0 eV along -$M\Gamma M$ and (b) near E$_{F}$ along
$\Gamma KM$ with different $d$ values. The red line in (b) is
drawn to indicate the Dirac point. The inset is the enlarged bands
near the Dirac point. }
\end{figure}

The energy bands of Gr/Au in the configuration (II), also with a
fine lattice match, are given in Fig. 2, where different
separation (d) between graphene and the Au substrate is
considered. The $\pi $ bands of graphene are accompanied by a gold
SS at = -7.3 eV\cite{liu}, labeled as 'Au SS' in Fig. 2(a). This
Au SS is actually localized at the vacuum side of the Au slab and
hence shows no change for different d in Fig. 2(a). The
corresponding Au SSs at the graphene side move down quickly with
the decrease of d due to the effect of graphene. At the
equilibrium geometry (d$_{eq}$=3.3 \r{A}), the SO splitting of
graphene $\pi $ bands at 0.3 $\Gamma M$ is about 21 meV, much
larger than that on the Ni substrate. For Gr/Au in the
configuration (II), the \textit{A} and \textit{B} sublattices
experience the same environment again. Thus, the graphene layer
almost restores its unique electronic structure: the Dirac cone
near E$_{F}$. Charge transfer between graphene and the metal and
therefore the energy position of the Dirac cone can be adjusted
through changing d.\cite{gio} At the equilibrium separation, the
Dirac point (Fig.2(b)) is very close to E$_{F}$, indicating almost
no charge transfer between graphene and Au. When the separation
expands/shrinks, the Dirac point moves upward/downward, revealing
net charge transfer between graphene and Au occurs. The SO
splitting of graphene, however, reduces both ways, independent of
the enhancement of effective electric field in the interface. This
contradicts to the effective electric field model proposed in
Refs.\cite{ded,rad} for the explanation of enhanced SO effect in
the systems. The inset in Fig.2(b) shows the SO splitting of the
Dirac point. Both the electron and hole bands show SO splitting of
about 5 meV, close to the value (13 meV) given in recent
experiment.\cite{va} The splitting at the Dirac point is induced
by the metal, as intrinsic SO interaction at this point was
predicted to be zero. \cite{ya}

\begin{figure}[tbp]
\resizebox{7.2cm}{!}{\includegraphics*[25,344][325,682]{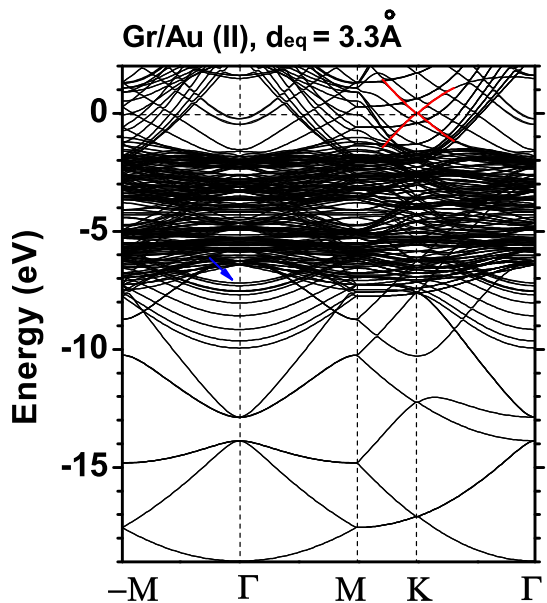}}
\caption{(Color online) The energy bands of Gr/Au(II) with SO
interaction in large energy range. The blue arrow indicates the
graphene $\protect\pi $ bands. The red lines indicate the Dirac
point. }
\end{figure}

\begin{table}[tbp]
\caption{The components of wave functions for graphene
$\protect\pi $ bands at $\Gamma $ for the considered systems at
equilibrium distances. The values are in the scale of the graphene
$p_{z}$ state. The bold expresses the more stable
configuration. The SO splitting (in meV) of graphene $\pi$ bands in the configuration (I) was given at $k=0.25$ $%
\Gamma M$, while in the configuration (II), at $k=0.5$ $\Gamma M$
due to the double
lattice vectors.}\tabcolsep0.5 mm 
\par
\begin{center}
\begin{tabular}{cccccc}
\hline\hline
~ & Ni (\textbf{I}) & Au (\textbf{II}) & Au (I) & Ag (\textbf{II}) & Ag (I)
\\
~ & \ (N=13) \  & \ (N=9) \  & \ (N=12) \  & (N=9) & \ (N=12) \  \\ \hline
$\ $Gr-$p_{z}$ state & 1.0 & 1.0 & 1.0 & 1.0 & 1.0 \\
$\ $M-$s$ state & 0.8 & 1.3 & 0.1 & 0.5 & 0.7 \\
$\ $M-$d_{z^{2}}$ state & 0.0 & 2.4 & 1.3 & 0.0 & 3.7 \\
$\ \text{SO splitting}$ & 6.3 & 32.2 & 89.0 & 1.6 & 36.9
\\ \hline\hline
\end{tabular}%
\end{center}
\end{table}

Figure 3 shows the energy bands in large energy range for Gr/Au
with configuration (II) at equilibrium distance. The two sets of
parabolas at about -0.2 and -0.4 eV around the $\Gamma $ point are
the Au (111) SSs at the two sides of the Au slab, with and without
graphene, respectively. They are degenerate at -0.4 eV for the
pure Au (111) film\cite{nico}, but now the SS in the side with the
adsorption of graphene is pushed up slightly. In comparison with
Fig. 1(b), the bulk Au 5$d$ bands around $\Gamma $ point (about
-2$\sim $-7 eV) are deeper in energy than Ni 3$d$ bands. This
bestows the probability for Au 5$d$ to strongly interact with
graphene $\pi $ bands, which are in a big interstice formed mainly
by Au 5$d$ states. The situation is very different from that in
Fig. 1(b), where the graphene $\pi $ and Ni 3$d$ states are well
separated in energy. To understand the different SO splittings in
Ni and Au cases, wave function compositions of the $\pi $ bands at
$\Gamma $ are analyzed quantitatively. As shown in Table I, the
ratio of C 2$p_{z}$:Ni 4$s$ in Gr/Ni (I) $\approx $ 1.0:0.8; no Ni
3$d$ is involved. Since Ni 4$s$ states do not contribute to the SO
effect, the enhancement of Rashba splitting in Gr/Ni only results
from the asymmetric potential distribution
in the two sides of graphene. In contrast, contribution from the Au $%
d_{z^{2}}$ state is obvious for Gr/Au (II); the ratio of C 2$p_{z}$:Au 6$s$%
:Au 5$d$ is about 1.0:1.3:2.4. Therefore, strong hybridization of the metal $%
d_{z^{2}}$ with graphene $p_{z}$ is a key factor to produce large
SO splitting in graphene $\pi $ bands.

\begin{figure}[tbp]
\resizebox{6cm}{!}{\includegraphics*[97,304][359,681]{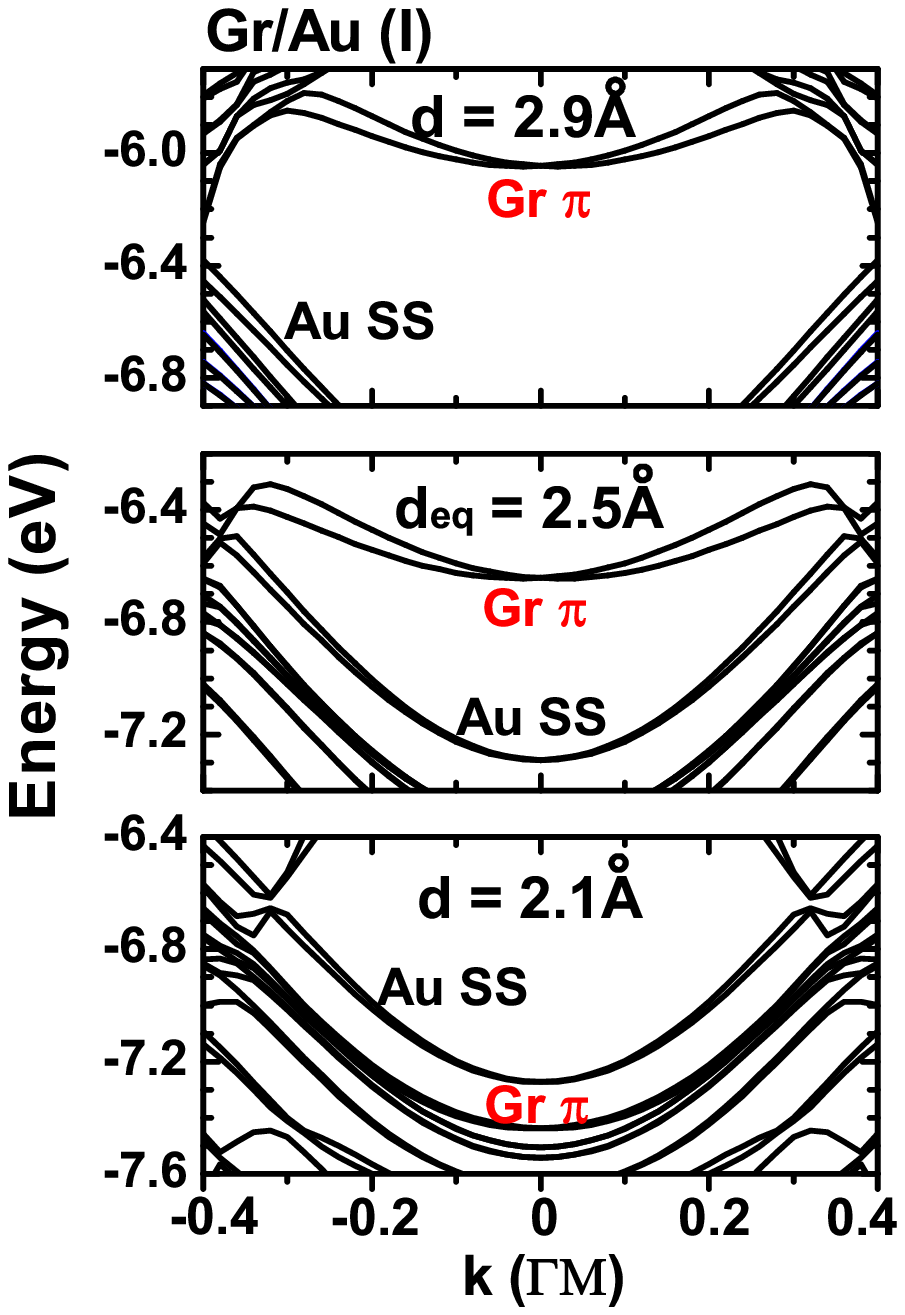}} %
\resizebox{1.8cm}{!}{\includegraphics*[95,281][188,714]{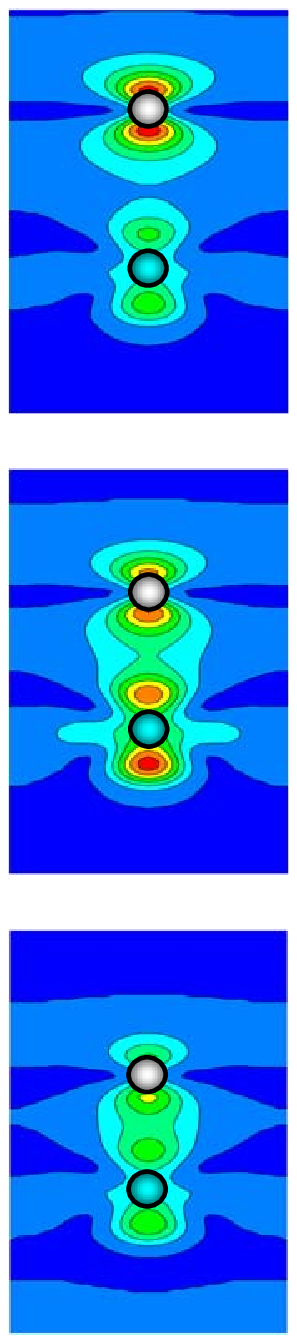}}
\caption{(Color online) Left: The bands of Gr/Au (I) along
-$M\Gamma M$\ with different d values. Right: The corresponding
charge density of graphene $\protect\pi $ bands at the $\Gamma $
point. The density is plotted in the plane perpendicular to the
interface, indicated by the dash line in Fig.1(a). The upper and
lower atoms are a carbon and a gold atom, respectively.}
\end{figure}

For Gr/Au, configuration (I) is less stable than configuration
(II) since the C-C bonds in graphene have to be stretched by 17\%
to match the structure of the Au (111) substrate. This stretching
gives rise to a much shorter equilibrium distance (2.5 \r{A}) in
the interface, and thus allows us to understand the effects of
adjusting the lateral and vertical distances. The whole $\pi $
bands of graphene in the configuration (I) thus disperse less; and
shift upward as well, hopefully causing more effective
hybridization between C 2$p_{z}$ and Au 5$d$ states. Figure 4
gives the SO splitting of the graphene $\pi $ bands in Gr/Au (I)
at d$_{eq}$ to be close to 100 meV, 3 times larger than that in
Au (II) case. In Gr/Au (I), the Au 6$s$ component decreases meanwhile the C 2$%
p_{z}$ and the Au 5$d_{z^{2}}$ more effectively hybridize to each
other (see Table I), as expected. Similar to the trend in Fig.
2(a) either, increasing or decreasing d cause a decrease of the SO
splitting in Fig. 4. This trend can be rationalized by using the
real-space charge densities of the graphene $\pi $ bands at
$\Gamma $ in the right panels of Fig. 4. At the equilibrium
distance, a very obvious interaction between graphene and metal
and asymmetric charge distribution above and below the graphene
plane are observed, corresponding to large SO splitting. When d
becomes longer or
shorter than d$_{eq}$, the effective mixing between C $p_{z}$ and Au $%
d_{z^{2}}$ is weakened.

Table I also contains the results of Gr/Ag (I) and (II). Again,
the SO splitting in the configuration (II) is much less than that
in (I) due to weakened hybridization. While the graphene $p_{z}$
state has strong interaction with the Ag $d_{z^{2}}$ state in the
configuration (I), only Ag $s$ state is involved in the more
stable Gr/Ag (II), similar to Ni. Therefore, heavy metals may not
always produce large SO splitting in graphene.

\begin{table}[tbp]
\caption{The SO splitting of $\protect\pi $ bands of graphene on
different metal substrates with N=1 and 6, respectively. The value
is given at $k=0.25$ or $0.5$ $\Gamma M$, as stated in Table I.
The bold indicates the more stable configuration.}
\begin{center}
\begin{tabular}{cccccc}
\hline\hline meV & \ \ \ Ni \textbf{(I) \ \ } & \ Au (\textbf{II})
\  & \ Au (I) \  & \ Ag (\textbf{II}) \  & $\ \text{Ag (I) \ }$ \\
\hline
\ \ N = $1$ \  & $6.7$ & $11.2$ & $35.4$ \  & $2.6$ & $6.7$ \  \\
N = $6$ & $6.3$ & $33.4$ & $88.0$ \  & $2.4$ & $41.8$ \  \\
\hline\hline
\end{tabular}%
\end{center}
\end{table}

Finally, we also explored the effect of the thickness of metal
films. The SO splitting of graphene $\pi $ bands for Gr/Au (I) and
(II) and Gr/Ag (I) with N=6 in Table II are almost the same as the
values listed in Table I, respectively. For the rest two cases:
Gr/Ni (I) and Gr/Ag (II), only one monolayer of metal substrate is
enough to give a saturated SO splitting of graphene. Since Au SSs
usually extend several atomic layers into the bulk\cite{liu}, a
few layers of metals are needed to obtain the saturated SO
splitting. Nevertheless, the SO splitting for graphene on Au mono-
and bi-layer films is already large, explaining why one Au atomic
layer intercalated between
graphene and Ni (111) can cause a substantial Rashba effect in the experiment.%
\cite{va}

In conclusion, we investigated what determines the SO splitting of graphene
bands on Ni, Au, or Ag (111) substrates through first-principles
calculations. While the Rashba splitting for Gr/Au is sizeable, the effect
of Ni is very limited. The hybridization between graphene $p_{z}$ and metal $%
d_{z^{2}}$ states is identified as the chief factor for the enhancement of
the SO effect in graphene. This requires not only large SO strength from
metal atoms, but also effective overlap of metal d and graphene states in
energy. A few atomic layers of metals are sufficient to produce saturated
strong Rashba splitting in graphene. Our findings point out a direction for
the manipulation of SO strength in graphene that is needed for the
development of spintronic materials and devices.

The work was supported by National Natural Science Foundation of China
(No.10674027), 973 project (No.2006CB921300), Fudan High-end Computing
Center, and Chemistry and Materials Research Division (MRSEC program) of NSF
in USA. Work at UCI was supported by DOE grant DE-FG02-05ER46237 and
computing time at NERSC.


\end{document}